\documentclass[aps,prl,superscriptaddress,showpacs]{revtex4}
\usepackage{epsfig}
\usepackage{amsmath}

\newcommand{\be}{\begin{equation}}
\newcommand{\ee}{\end{equation}}
\newcommand{\beq}{\begin{eqnarray}}
\newcommand{\eeq}{\end{eqnarray}}
\newcommand{\mbf}{\mathbf}
\newcommand{\ds}{\displaystyle}

\usepackage{pstricks}
\newif\ifcolor
\colortrue
\ifcolor
  \newrgbcolor{red}{0.9 0 0}
  \newrgbcolor{green}{0 0.5 0.}
  \newrgbcolor{blue}{0 0 1}
  \newrgbcolor{mag}{0.8 0 0.8}
\else
  \newrgbcolor{red}{0 0 0}
  \newrgbcolor{green}{0 0 0}
  \newrgbcolor{blue}{0 0 0}
  \newrgbcolor{mag}{0 0 0}
\fi

\begin{document}
\title{Estimating Granger causality from Fourier and wavelet transforms of time series data}
\author{Mukeshwar Dhamala}
\affiliation{Department of Physics and Astronomy, Brains and Behavior Program, Georgia State University, Atlanta, GA 30303, USA}
\author{Govindan Rangarajan}
\affiliation{Department of Mathematics, Indian Institute of Science, Bangalore 560012, India}
\author{Mingzhou Ding}
\affiliation{Department of Biomedical Engineering, University of Florida, Gainesville, FL 33611, USA}
\date{\today}
\begin{abstract}
Experiments in many fields of science and engineering yield data in
the form of time series. The Fourier and wavelet transform-based
nonparametric methods are used widely to study the spectral
characteristics of these time series data. Here, we extend the
framework of nonparametric spectral methods to include the
estimation of Granger causality spectra for assessing directional
influences. We illustrate the utility of the proposed methods using
synthetic data from network models consisting of interacting
dynamical systems. [{\green{\bf Physical Review Letters,  in press}}].
\end{abstract}
\maketitle
Extracting information flow in networks of coupled
dynamical systems from the time series measurements of their
activity is of great interest in physical, biological and social
sciences. Such knowledge holds the key to the understanding of
phenomena ranging from turbulent fluids to interacting genes and
proteins to networks of neural ensembles. Granger causality
\cite{Granger:1969} has emerged in recent years as a leading
statistical technique for accomplishing this goal. The definition of
Granger causality \cite{Granger:1969} is based on the theory of
linear prediction \cite{Wiener:1956} and its original estimation
framework requires autoregressive (AR) modeling of time series data
\cite{Granger:1969, Geweke:1982}. Such parametric Granger causality
and associated spectral decompositions have been applied in a wide
variety of fields including condensed matter physics
\cite{Ganapathy:2007},
neuroscience~\cite{Kaminski:2001,Brovelli:2004,Goebel:2003,Seth:2007},
genetics \cite{Mukhopadhyay:2007}, climate science
\cite{Kaufmann:1997,Mosedale:2006}, and economics
\cite{Granger:1969,Hiemstra:1994}. However, the parametric modeling
methods often encounter difficulties such as uncertainty in model
parameters and inability to fit data with complex spectral contents
\cite{Mitra:1999}. On the other hand, the Fourier and wavelet
transform-based nonparametric spectral methods are known to be free
from such difficulties \cite{Mitra:1999} and have been used
extensively in the analysis of univariate and multivariate
experimental time series \cite{SpectralMethods, Percival:2000}. A
weakness of the current nonparametric framework is that it lacks the
ability for estimating Granger causality. In this Letter, we
overcome this weakness by proposing a nonparametric approach to
estimate Granger causality directly from Fourier and wavelet
transforms of data, eliminating the need of explicit AR modeling.
Time-domain Granger causality can be obtained by integrating the
corresponding spectral representation over frequency
\cite{Geweke:1982}. Below, we present the theory and apply it to
simulated time series.

{\it Granger causality: the parametric estimation approach}. Granger
causality \cite{Granger:1969} is a measure of causal or directional
influence from one time series to another and is based on linear
predictions of time series. Consider two simultaneously recorded
time series: $X_1:$ $x_1(1),x_1(2),...,x_1(t),...$;  $X_2:$
$x_2(1),x_2(2),...,x_2(t),...$ from two stationary stochastic
processes ($X_1, X_2$).  Now, using AR representations, we construct
bivariate linear prediction models for $x_1(t)$ and $x_2(t)$:
\begin{eqnarray}
\label{eq:Eq1}
x_1(t)& = &\sum_{j=1}^{\infty}b_{11,j}x_1(t-j)+ \sum_{j=1}^{\infty}b_{12,j}x_2(t-j)+\epsilon_{1|2}(t) \\
\label{eq:Eq2}
x_2(t)& = &\sum_{j=1}^{\infty}b_{21,j}x_1(t-j)+ \sum_{j=1}^{\infty}b_{22,j}x_2(t-j)+\epsilon_{2|1}(t)
\end{eqnarray}
along with the univariate models:  $x_1(t) =
\sum_{j=1}^{\infty}\alpha_jx_1(t-j)+\epsilon_{1}(t)$ and $x_2(t) =
\sum_{j=1}^{\infty}\beta_jx_2(t-j)+\epsilon_{2}(t)$. Here,
$\epsilon$'s are the prediction errors. If ${\rm
var}(\epsilon_{1|2}(t))<{\rm var}(\epsilon_{1}(t))$ in some suitable
statistical sense, then $X_2$ is said to have a causal influence on
$X_1$. Similarly, if ${\rm var}(\epsilon_{2|1}(t))<{\rm
var}(\epsilon_{2}(t))$, then there is a causal influence from $X_1$
to $X_2$.  These causal influences are quantified in time domain
\cite{Geweke:1982} by $F_{j\rightarrow i} = \ln \ds\frac{{\rm
var}(\epsilon_{i}(t))}{{\rm var}(\epsilon_{i|j}(t))}$, where $i=1,2$
and $j=2,1$.

Experimental processes are often rich in oscillatory content,
lending themselves naturally to spectral analysis. The spectral
decomposition of Granger's time-domain causality was proposed by
Geweke in 1982 \cite{Geweke:1982}. To derive
the frequency-domain Granger causality, we start with Eq.~(\ref{eq:Eq1}-\ref{eq:Eq2}).
We rewrite these equations in a matrix form with a lag operator
$L$: $Lx (t)=x(t-1)$ as
\begin{eqnarray}
\label{eq:Eq3}
\left(\begin{array}{cc}
  b_{11}(L) \ \ & b_{12}(L) \\
  b_{21}(L) \ \ & b_{22}(L) \\
\end{array}
\right)\left(\begin{array}{c}
  x_1(t) \\ x_2(t) \\
\end{array}\right)
&=&
\left(\begin{array}{c}\epsilon_{1|2} \\
  \epsilon_{2|1} \\
\end{array}\right),
\end{eqnarray}
where $b_{ij}(L)=\sum_{k=0}^{\infty} b_{ij,k} L^k$ with $b_{ij,0} =\delta_{ij}$ (the Kronecker delta function). The
covariance matrix of the noise terms is $\mbf \Sigma = \left(\begin{array}{cc}\Sigma_{11} \ \ &
\Sigma_{12}\\\Sigma_{21}\ \ & \Sigma_{22} \\\end{array}\right)$
where $\Sigma_{11}={\rm var}(\epsilon_{1|2})$, $\Sigma_{12}=\Sigma_{21}={\rm cov}(\epsilon_{1|2},
\epsilon_{2|1})$, and $\Sigma_{22}={\rm var}(\epsilon_{2|1})$.
Fourier transforming Eq.~(\ref{eq:Eq3}) yields
\begin{eqnarray}
\label{eq:Eq4}
\left(\begin{array}{cc}
  B_{11}(f) \ \ & B_{12}(f) \\
  B_{21}(f) \ \ & B_{22}(f) \\
\end{array}\right)
\left(\begin{array}{c}
  X_1(f) \\
  X_2(f) \\
\end{array}\right) =\left(
\begin{array}{c}
  E_1(f) \\
  E_2(f) \\
\end{array}\right),
\end{eqnarray}
where the components of the coefficient matrix $[B_{ij}(f)]$ are
$B_{lm}(f)=\delta_{lm}-\sum_{k=1}^\infty b_{lm,k}e^{-i2\pi fk}$.
In terms of transfer function matrix ($\mbf H(f) = [B_{ij}(f)]^{-1}$), Eq.~(\ref{eq:Eq4}) becomes
\begin{eqnarray}
\label{eq:Eq5}
\left(\begin{array}{c}
  X_1(f) \\
  X_2(f) \\
\end{array}
\right) =  \left(\begin{array}{cc}
  H_{11}(f) \ \ & H_{12}(f) \\
  H_{21}(f) \ \ & H_{22}(f) \\
\end{array}\right)\left(
\begin{array}{c}
  E_1(f) \\
  E_2(f) \\
\end{array}\right).
\end{eqnarray}
Then, the spectral density matrix $\mbf S(f)$ is given by
\begin{eqnarray}
\label{eq:Eq6}
\mbf S(f) &=&\mbf H(f)\mbf\Sigma\mbf H^*(f),
\end{eqnarray}
where $*$ denotes matrix adjoint. To examine the causal influence
from $X_2$ to $X_1$, one needs to look at the auto-spectrum of
$x_1(t)$-series, which is $S_{11}(f)=H_{11}\Sigma_{11}H_{11}^* +
2\Sigma_{12} {\rm Re} (H_{11} H_{12}^*) + H_{12}
\Sigma_{22}H_{12}^*$. Here, because of the cross-terms in this
expression for $S_{11}$, the causal power contribution is not
obvious. Geweke \cite{Geweke:1982} introduced a transformation that
eliminates the cross terms and makes an intrinsic power term and a
causal power term identifiable. For $X_1$-process, this
transformation is achieved by left-multiplying Eq.~(\ref{eq:Eq4}) on
both sides with $\left( \begin{array}{cc}
1 & 0 \\
-\Sigma_{12}/\Sigma_{11} & 1\\
\end{array}\right)$, which yields:
\begin{eqnarray}
\label{eq:Eq7}
\left(
\begin{array}{cc}
  B_{11}(f) & B_{12}(f) \\
  \tilde{B}_{21}(f) & \tilde{B}_{22}(f) \\
\end{array}
\right) \left(\begin{array}{c}
  X_1(f) \\ X_2(f) \\
\end{array}
\right)= \left(\begin{array}{c}
  E_1(f) \\\tilde{E} _2(f)\\
\end{array}
\right),
\end{eqnarray}
where $\tilde{B}_{21}(f)=B_{21}(f)-\ds{\frac{\Sigma_{12}}{\Sigma_{11}}}
B_{11}(f)$, $\tilde{B}_{22}(f)=B_{22}(f)-\ds{\frac{\Sigma_{12}}{\Sigma_{11}}} B_{12}(f)$,
and $\tilde{E}_2(f)=E_2(f)-\ds{\frac{\Sigma_{12}}{\Sigma_{11}}} E_1(f)$.  The elements of the new
transfer function $\tilde{\mbf H}(f)$ then become $\tilde{H}_{11}(f)=H_{11}(f)+\ds{\frac{\Sigma_{12}}{\Sigma_{11}}}H_{12}(f)$,
$\tilde{H}_{12}(f)=H_{12}(f)$, $\tilde{H}_{21}(f)=H_{21}(f)+\ds{\frac{\Sigma_{12}}{\Sigma_{12}}}H_{11}(f)$, and
$\tilde{H}_{22}(f)=H_{22}(f)$. Here, $\rm{cov}(E_1,\tilde{E}_2)=0$ and the new variance of $
x_2(t)$ is $\tilde{\Sigma}_{22}=\Sigma_{22}-\ds{\frac{\Sigma_{12}^2}{\Sigma_{11}}}$.  Now, the auto-spectrum of $x_1(t)$ is decomposed
into two obvious parts:
$S_{11}(f)=\tilde{H}_{11}(f)\Sigma_{11}\tilde{H}_{11}^*(f) + H_{12}(f)\tilde{\Sigma}_{22} H_{12}^*(f)$,
where the first term accounts for the intrinsic power of $x_1(t)$ and the second term for causal power due to the influence
from $X_2$ to $X_1$.  Since Granger causality is the natural logarithm of the ratio of total power to intrinsic power \cite{Geweke:1982},
causality from $X_2$ to $X_1$ (or, $2$ to $1$) at frequency $f$ is
\begin{eqnarray}
\label{eq:Eq8}
I_{2\rightarrow 1}(f)&=&\ln\frac{S_{11}(f)}{S_{11}(f)-\left(\Sigma_{22}-\ds{\frac{\Sigma_{12}^2}{\Sigma_{11}}}\right)|H_{12}(f)|^2},
\end{eqnarray}
using the expressions for $S_{11}$ and $\tilde{\Sigma}_{22}$ obtained after the transformation. Next, by
taking the transformation matrix as $\left( \begin{array}{cc}1 &
-\Sigma_{12}/\Sigma_{22}\\0& 1\\\end{array}\right)$ and performing
the same analysis, one can get Granger causality $I_{1\rightarrow
2}(f)$ from $X_1$ to $X_2$, the expression for which can be obtained
just by exchanging subscripts $1$ and $2$ in Eq.~(\ref{eq:Eq8}).
Geweke\cite{Geweke:1982} showed that the time-domain measure is
theoretically related to the frequency-domain measure as
$F_{2\rightarrow 1}\le\ds{\frac{1}{2\pi}}\int_{-\pi}^\pi
I_{2\rightarrow 1}(f) df$, but for all processes of practical
interest, the equality holds.

From the above discussion, it is clear that the estimation of
frequency-domain Granger causality requires noise covariance and
transfer function which are obtained as part of the AR data
modeling. The mathematics behind this parametric approach to obtain
these quantities is well-established. However, for nonparametric
methods the current estimation framework does not contain
provisions for computing these quantities. Moreover, the parametric
estimation method from finite data can often produce erroneous
results if the series in Eq.~(\ref{eq:Eq1}-\ref{eq:Eq2}) are not truncated to proper model
orders. There are criteria \cite{AIC_BIC} for choosing proper AR
model order, but these criteria cannot always be satisfied. In
addition, AR modeling approach does not always capture all the
spectral features \cite{Mitra:1999}. To overcome these difficulties,
we propose a nonparametric estimation approach, in which we derive,
based on Fourier and wavelet transforms of time series data, noise
covariance and transfer function to be used in Geweke's formulae
such as Eq.~(\ref{eq:Eq8}) for Granger causality estimates.

{\it Granger causality: the nonparametric estimation approach}.  In
the nonparametric approach, spectral density matrices are estimated
by using direct Fourier and wavelet transforms of time series data.
These matrices then undergo spectral density matrix
factorization \cite{Sayed:2001,Wilson:1972_1978} and Geweke's
variance decomposition \cite{Geweke:1982} for the estimation of
Granger causality. To explain this approach, let us consider a
bivariate process with time series $x_1(t)$ and $x_2(t)$,  their
Fourier transforms: $X_1(f)$ and $X_2(f)$, and wavelet transforms:
$W_{X_1}(t,f)$ and $W_{X_2}(t,f)$. Then, the spectral matrix ${\bf S}$ is
defined as:
${\bf S} = \left(
\begin{array}{cc}
  S_{11} & S_{12}\\
  S_{21} & S_{22}\\
\end{array}\right)$,
where, in the Fourier-based method, $S_{lm} =
\left<X_l(f)X_m(f)^*\right>$, and, in the wavelet method, $S_{lm}
=\left<W_{X_l}(t,f)W_{X_m}(t,f)^*\right>$.  Here, $l = 1,~2$,
$m=1,~2$, and $\left<.\right>$ is averaging over multiple
realizations. Smoother Fourier-based spectral density with reduced
estimation bias can be obtained by using the multitaper technique
\cite{Thomson:1982,Mitra:1999}, which involves the use of discrete
spheroidal sequences (DPSS) \cite{Slepian:1961}. The continuous
wavelet transform $W_{X_l}(t,s)$ at time $t$ and scale $s$ is
computed by the convolution of time series $x_l$ with a scaled and
translated version of a prototype wavelet $\Psi(\eta)$ that
satisfies zero-mean and unity square-norm conditions
\cite{Daubechies:1990,Torrence:1998}:
$W_{X_l}(t,s)=\ds\frac{1}{\sqrt{s}}\int_{-\infty}^{\infty}d\eta\Psi^*\left(\ds\frac{\eta-t}{s}\right)x_l(\eta)$.
Using the relationship between $s$ and $f$ for a given prototype
wavelet, such as the Morlet wavelet
\cite{Morlet:1984,Torrence:1998}, one can transform $W_{X_l}(t,s)$
into $W_{X_l}(t,f)$.  The wavelet transform at $f = 0$ can be
obtained by a numerical extrapolation. ${\bf S}(f)$ or ${\bf
S}(t,f)$ thus formed is a square matrix that can be defined in the
interval $[-\pi, \pi]$ and, for all processes of practical interest,
satifies the following properties: (i) $S(\theta)$ is Hermitian,
nonnegative, and $S(-\theta)=S^T(\theta)$, where $\theta = 2\pi f$
and $^T$ denotes matrix transpose, and (ii) $S(\theta)$  is
integrable and has a Fourier series expansion: $S(\theta) =
\sum_{k=-\infty}^{\infty}\gamma_ke^{ik\theta}$, where the covariance
sequence $\{\gamma_k\}_{-\infty}^\infty$ is formed by
$\gamma_k=(1/2\pi)\int_{-\pi}^{\pi}S(\theta)e^{-ik\theta}d\theta$.

According to the factorization theorem
\cite{Masani:1966,Wilson:1972_1978},  the spectral density matrix
${\bf S}$ as defined above can be factored~\cite{Wilson_algorithm}
into a set of unique minimum-phase functions:
\begin{eqnarray}
\label{eq:Eq9}
{\mbf S} &= &{\mbf \psi}{\mbf \psi}^*,
\end{eqnarray}
where $^*$ denotes matrix adjoint, ${\mbf \psi}(e^{i\theta}) =
\sum_{k=0}^\infty{\mbf A}_ke^{ik\theta}$ is defined on the unit
circle $\{|z|=1\}$, and ${\mbf A}_k=(1/2\pi)\int_{-\pi}^{\pi}{\mbf
\psi}(e^{i\theta})e^{-ik\theta}d\theta$. Moreover, ${\mbf \psi}$ can
be holomorphically extended~\cite{Wilson_algorithm} to the inner
disk $\{|z|<1\}$ as ${\mbf \psi}(z)=\sum_{k=0}^{\infty} A_k z^k$
where $\mbf \psi(0) = \mbf A_0$, a real, upper triangular matrix
with positive diagonal elements. Similarly $S$ and $H$ can be
defined as functions of $z$ with $H(0)=I$. Comparing the right hand
sides of Eqs~(\ref{eq:Eq6}) and (\ref{eq:Eq9}) at $z=0$ we get
\begin{eqnarray}
\label{eq:Eq10}
\mbf \Sigma & = & \mbf A_0 \mbf A_0^T.
\end{eqnarray}
Rewriting Eq.~(\ref{eq:Eq9}) as $\mbf S = \mbf {\psi A_0^{-1}A_0A^TA_0^{-T}\psi^*}$
and comparing with Eqs~(\ref{eq:Eq6}) and (\ref{eq:Eq10}), we arrive at the expression for
the transfer function:
\begin{eqnarray}
\label{eq:Eq11}
\mbf H &=& \mbf \psi A_0^{-1}
\end{eqnarray}
\begin{figure}
\epsfig{figure=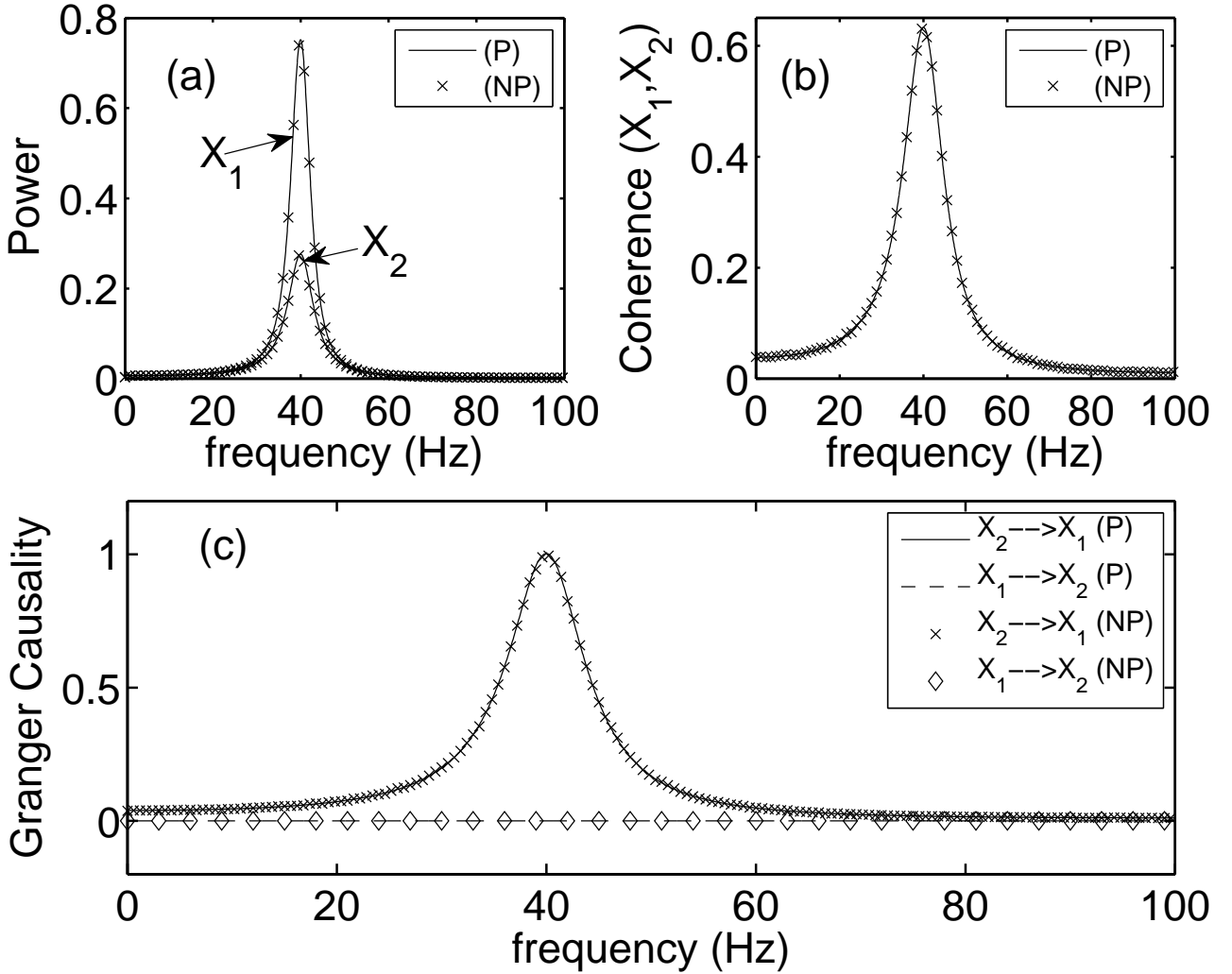,width=1\linewidth} \caption{(a) Power,
(b) coherence, and (c) Granger causality spectra from both Fourier
transform-based nonparametric (NP) and parametric (P) methods. There
is an excellent agreement between NP and P estimates. }
\label{fig:fig1}
\end{figure}
\begin{figure}
\epsfig{figure=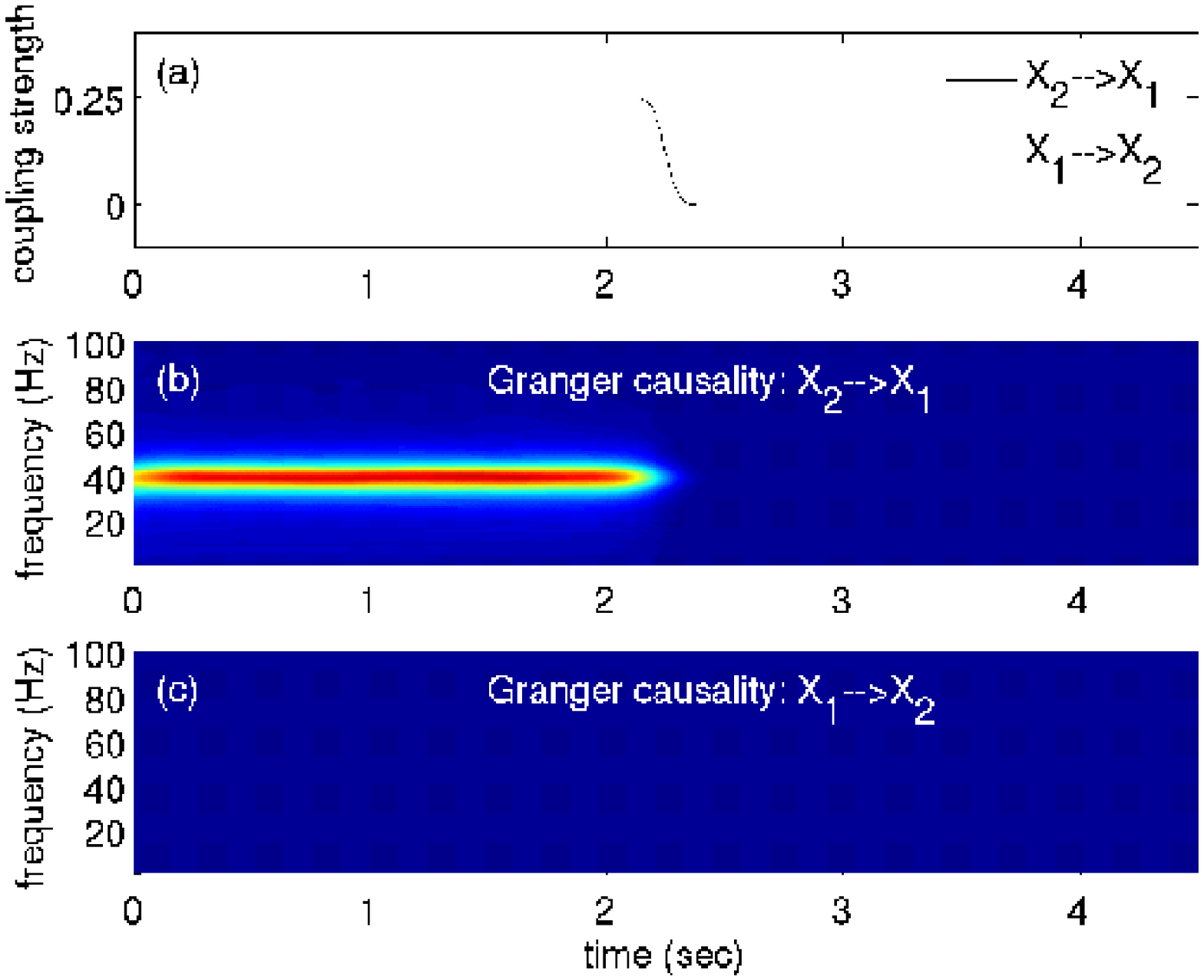,width=1.\linewidth}
\caption{(color online). Wavelet-based Granger causality: time-frequency representation of causality. Fig (a): temporal
structures of couplings constructed in the network model: the coupling of $X_2$ with $X_1$ stays $0.25$ during time $[0, 2]$ sec, slowly
changes to $0$  during $[2, 2.25]$ sec, and stays $0$ during time $> 2.25$ sec, whereas the coupling of $X_1$ with $X_2$ is $0$ throughout.
The slow transition in the middle is modeled by the tangent of a hyperbolic function. Fig (b, c): time-frequency maps of Granger causality
correctly represent the temporal structures of couplings as in Fig (a) for the network model.}
\label{fig:fig2}
\end{figure}
Now, by substituting the specific elements of the noise covariance
and transfer function from Eq.~(\ref{eq:Eq10}) and (\ref{eq:Eq11})
into Eq.~(\ref{eq:Eq8}), one can estimate pairwise Granger causality
spectra.  In case of the wavelet, these calculations are repeated
along the time axis for each time point to get the time-frequency
representation of Granger causality.

{\it Numerical examples}.  We consider network models with two
autoregressive processes $X_1$ and $X_2$ as nodes where $X_1(t) =
0.55 X_1(t-1) - 0.8 X_1(t-2)+ C(t) X_2(t-1) + \epsilon (t) $ and
$X_2(t) = 0.55 X_2(t-1) - 0.8 X_2(t-2)+ \xi (t)$. Here, $t$ is a
time index, $\epsilon(t)$ and $\xi(t)$ are independent white noise
processes with zero means and unit variances, $C(t)$ is the coupling
strength, and the sampling rate is considered to be $200$ Hz. By
construction, only $X_2$ has a causal influence on $X_1$.  First, we
fix $C(t)$ at $0.2~\forall~t$, generate dataset of 5000 realizations
(trials), each consisting of 5000 data points, and apply the
Fourier-based nonparametric method.  The power spectra of $X_1$ and
$X_2$ (Fig. 1(a)) and coherence spectra between $X_1$ and $X_2$
(Fig. 1 (b)) show $40$ Hz peaks. Figure 1 (c) shows the Granger
causality spectra. Here, both the nonparametric (NP) and parametric
(P) approaches yield identical results, recovering correctly the
underlying directional influences. Since the proper model order is
chosen here and the dataset is large, the parametric causality
estimates can be assumed to represent the theoretical values. Next,
we let the unidirectional coupling of $X_2$ to $X_1$ change in its strength
$C(t)$ over time as shown in Fig~2(a), generate $5000$ realizations,
each with $900$ time-points. Then, the wavelet spectra are computed
for all the trials using the Morlet wavelet (as used in
\cite{Torrence:1998}). The average wavelet spectra are obtained by
averaging over these individual spectra. The average spectra at a
time point is subsequently factored, and ${\bf H}$ and ${\bf \Sigma}$
are obtained and used in Eq.~(\ref{eq:Eq8}) to obtain Granger
causality spectra.  By repeating these calculations along time axis,
one gets the complete time-frequency maps of Granger causality (Fig.
2(b-c)), which also recovers the correct directional influences.
Granger causality magnitude increases with coupling strength.

Here, the proposed Granger causality techniques are tested on
datasets with a large number of long trials. However, these
techniques can also be used reliably with fewer trials. Increasing
the number of trials leads to spectral estimates with smaller
variance. A single, sufficiently long stationary time series can be
broken into smaller segments, each of which can be treated as a
distinct trial. The use of multitaper \cite{Mitra:1999} and
multiwavelet \cite{Brittain:2007} techniques can yield better
estimates of Granger causality in case of a dataset with shorter
length and fewer trials. See Supplementary Material \cite{Supp}
for additional numerical examples and applications to brain signals.

{\it Conclusion}. Granger causality is a key technique for
assessing causal relations and information flow among
simultaneous time series. Its traditional parametric estimation framework
often suffers from uncertainty in model order selection and inability to fully
account for complex spectral features. We develop a nonparametric
approach based on the direct Fourier and wavelet transforms of data that
eliminates the need of parametric data modeling and extends the capability of
Fourier and wavelet-based suites of nonparametric spectral tools. It
is expected that the integration of the proposed method into
existing laboratory analysis routines will provide the basis for
gaining deeper insights into the organization of dynamical networks
arising in many fields of science and engineering
\cite{Other_techniques}.

We thank G. T. Wilson for useful email communications. This work was
supported by NIH Grant MH71620. GR was also supported by grants from
DRDO, DST (SR/S4/MS:419/07) and UGC (under SAP-Phase IV).

\end{document}